\pdfoutput=1

\documentclass[11pt]{article}

\usepackage{acl}

\usepackage{times}
\usepackage{latexsym}
\usepackage{makecell}

\usepackage[T1]{fontenc}

\usepackage[utf8]{inputenc}

\usepackage{inconsolata}

\usepackage{graphicx}
\usepackage{multirow}
\usepackage{url}
\usepackage{enumitem}
\usepackage{lscape}
\usepackage{rotating}

\title{A Report on Financial Regulations Challenge at COLING 2025 \\
\large FinNLP-FNP-LLMFinLegal-2025 Shared Task}

\author{
 \textbf{Keyi Wang\textsuperscript{1}},
 \textbf{Jaisal Patel\textsuperscript{2}},
 \textbf{Charlie Shen\textsuperscript{1}},
 \textbf{Daniel Kim\textsuperscript{2}},
 \textbf{Andy Zhu\textsuperscript{2}},
 \textbf{Alex Lin\textsuperscript{2}},
 \textbf{Luca Borella\textsuperscript{3}},
 \\
 \textbf{Cailean Osborne\textsuperscript{4}},
 \textbf{Matt White\textsuperscript{5}},
  \textbf{Steve Yang\textsuperscript{6}},
 \textbf{Kairong Xiao\textsuperscript{1}},
 \textbf{Xiao-Yang Liu Yanglet\textsuperscript{1,2}}
\\
 \textsuperscript{1}Columbia University,
 \textsuperscript{2}Rensselaer Polytechnic Institute,
 \textsuperscript{3}FINOS, Linux Foundation
 \\
 \textsuperscript{4}University of Oxford,
 \textsuperscript{5}PyTorch Foundation; GM of AI, Linux Foundation\\
 \textsuperscript{6}Stevens Institute of Technology
\\
\texttt{\{kw2914,cs4206,kx2139,XL2427\}@columbia.edu},
\texttt{\{patelj8,Kimd24,zhua6,lina\}@rpi.edu},
\\
\texttt{cailean.osborne@oii.ox.ac.uk},~ \texttt{luca.borella@finos.org }\\
\texttt{matt.white@linuxfoundation.org},~~\texttt{syang14@stevens.edu }
}

\begin{document}
\maketitle
\begin{abstract}

Financial large language models (FinLLMs) have been applied to various tasks in business, finance, accounting, and auditing. Complex financial regulations and standards are critical to financial services, which LLMs must comply with. However, FinLLMs' performance in understanding and interpreting financial regulations has rarely been studied. Therefore, we organize the Regulations Challenge\footnote{Website: \url{https://coling2025regulations.thefin.ai/home}}, a shared task at COLING FinNLP-FNP-LLMFinLegal-2025. It encourages the academic community to explore the strengths and limitations of popular LLMs. We create 9 novel tasks and corresponding question sets. In this paper, we provide an overview of these tasks and summarize participants’ approaches and results. We aim to raise awareness of FinLLMs' professional capability in financial regulations. 
\end{abstract}

\section{Introduction}

The financial industry follows strict regulations and industry standards to ensure market integrity, protect investor interests, and mitigate systemic risk ~\cite{markus2009regulations}. Large language models (LLMs) with remarkable capabilities in understanding and generating texts are promising tools to process and interpret financial regulations, with a rapidly growing number of LLMs available on Hugging Face Hub \cite{osborne2024ai}.

However, financial regulations and industry standards present unique challenges to the professional readiness of financial LLMs (FinLLMs). The complex regulatory framework and overlapping jurisdictions, such as the fragmented dual federal-state framework in the U.S., make the compliance process challenging ~\cite{labonte2023framework}. Financial regulation requires processing multimodal data \cite{liu2024MFFM}, including, but not limited to, legal texts, financial statements, mathematical formulas, tables, figures, and charts. 
Moreover, LLMs face issues with misinformation and hallucinations, where they generate inaccurate or seemingly plausible but fabricated information ~\cite{kang2023deficiency}. Such hallucinations or misinformation are unacceptable in deployment and can lead to regulatory violations, substantial monetary losses, and erosion of trust between companies and their customers \cite{roberts2023}. 

To evaluate LLMs' capabilities in financial regulations, we organize the \textbf{Regulations Challenge}, a shared task at COLING FinNLP-FNP-LLMFinLegal-2025. It aims to challenge the academic community to explore the strengths and limitations of LLMs in financial regulations and industry standards. We designed 9 novel tasks to evaluate LLMs in 5 areas: information retrieval, passing certificates, the Common Domain Model (CDM), the Model Openness Framework (MOF), and eXtensible Business Reporting Language (XBRL) analytics. For each task, we create a question set from diverse documents, such as regulatory filings and official documentation.

The remainder of this report is organized as follows. Section 2 describes the tasks and question sets. Section 3 discusses the participants' methods. Section 4 discusses their results. Section 5 concludes and recommends future research directions.

\section{Task and Dataset}
In this section, we present our nine novel tasks and the corresponding question sets. 

\begin{table*}[t!]
\centering
\begin{tabular}{p{1.5cm} p{2cm} p{11cm}}
\hline
\textbf{Category} & \textbf{Task}  & \textbf{Examples} \\ 
\hline\hline
\multirow{5}{5em}{Basic Capabilities} 
        & \multirow{2}{5em}{Abbreviation Recognition}  & IPO: Initial Public Offering \\
         & & ICO: Initial Coin Offering \\ 
\cline{2-3}
        & Definition Recognition   & Stakeholder: a party who has an interest and might be affected by the performance and outcome of an entity’s business, project, or enterprise. \\ 
\cline{2-3} 
        & Named Entity Recognition (NER)  & Regulation (EU) No 648/2012 of the European Parliament and of the Council of 4 July 2012 on OTC derivatives, central counterparties and trade repositories (“EMIR”) entered into force on 16 August 2012. \\ 
\cline{2-3} 
        & Question Answering      & How do Basel III regulations, including the FRTB, aim to enhance market stability?  \\ 
\cline{2-3} 
        & Link Retrieval           & Regulation (EU) 2019/834 -\url{https://eur-lex.europa.eu/eli/reg/2019/834/oj} \\ 
\hline
\multirow{1}{5em}{Passing Certificate}
        & Certificate Question  & Phil Jones, CFA,... is about to issue an unfavorable report on the company. His manager does not want him to state any adverse opinions... \\ 
\hline
\multirow{1}{5em}{CDM}
        & CDM  & How is the \texttt{TradeState} data type utilized to track changes in a trade's lifecycle in the Common Domain Model? \\ 
\hline
\multirow{1}{5em}{MOF}
        & Licenses  & What licenses are recommended for Model Parameters under the Model Openness Framework? \\ 
\hline
\multirow{1}{5em}{XBRL}
        & Analytics & What is the value of Walt Disney Company's total assets for the fiscal year ending in 2023? \\ 
\hline\hline
\end{tabular}
\caption{Overview of nine novel tasks with examples.}
\vspace{-0.2in}
\label{tab:tasks}
\end{table*}

\begin{table*}[t!]
  \centering
  \begin{tabular}{l p{4cm} c c p{5cm} }
    \hline
    \textbf{Question Sets} & \textbf{Domains}  & \textbf{Size} & \textbf{Metrics} & \textbf{Data Sources}\\
    \hline
    \multirow{5}{7em}{Abbreviation Dataset (\textbf{3562})} 
                            & EMIR & 115  & \multirow{5}{3.5em}{Accuracy} & ESMA       \\
                            & US financial laws & 76  &  &  SEC, FINRA       \\
                            & Federal Reserve & 44  & & Federal Reserve        \\
                            & Accounting and auditing & 29  &  &  FDIC, III, FASAB, SBOA       \\
                            & Stock tickers & 3298  & &   NYSE      \\
    \hline
    \multirow{4}{7em}{Definition Dataset (\textbf{193})} 
                            & EMIR & 50  & \multirow{4}{3.5em}{BertScore} & ESMA        \\
                            & Securities and Exchanges & 13  &  &   SEC      \\
                            & Federal Reserve &  100 &  & Federal Reserve        \\
                            & Accounting and auditing & 30  & &  FDIC, III, SBOA        \\
    \hline
    NER Dataset (\textbf{49}) & EMIR & 49 & F1 Score & EUR-LEX, ESMA \\
    \hline
    \multirow{3}{7em}{QA Dataset (\textbf{124})} 
                            & Securities and Exchanges & 19  & \multirow{3}{3.5em}{FActScore} &   SEC      \\
                            & Federal Reserve & 55 & &        Federal Reserve \\
                            & Accounting and auditing & 50 &  &  FDIC, III, SBOA, FASAB       \\
    \hline
    \multirow{4}{7em}{Link Retrieval Dataset (\textbf{183})} 
                            & EMIR & 100  & \multirow{4}{3.5em}{Accuracy} &  EUR-LEX, ESMA \\
                            & SEC & 18 & & SEC, eCFR \\
                            & FDIC & 49 & & FDIC, eCFR\\
                            & Federal Reserve & 16 &  & Federal Reserve, eCFR \\
    \hline
    \multirow{4}{7em}{Certificate Question Dataset (\textbf{346})} 
                            & CFA Level I & 90  & \multirow{4}{3.5em}{Accuracy}  &  CFA Level I (real + mock) \\
                            & CFA Level II & 77 &  & CFA Level II (real + mock) \\
                            & CFA Level III & 78 &  & CFA Level III (real + mock) \\
                            & CPA REG & 101 & & REG CPA mock exams \\
    \hline
    \multirow{6}{7em}{CDM Dataset (\textbf{126})} 
                            & Product model& 20  & \multirow{6}{3.5em}{FActScore} &  CDM documentation\\
                            & Event model & 20 &   & CDM documentation\\
                            & Legal agreements & 12 &  & CDM documentation\\
                            & Process model & 19 &  & CDM documentation \\
                            & General and Other & 9 &  & CDM documentation \\
                            & Implementation \& \hphantom{null} Deployment & 46 & & FAQ, CDM experts at FINOS \\                        
    \hline
    \multirow{3}{7em}{MOF Licenses Dataset (\textbf{161})} 
                            & License Abbreviations& 41  & Accuracy &  OSI website\\
                            & OSI Approval & 50 & Accuracy  & OSI website\\
                            & Detailed QA & 70 & FActScore & MOF paper\\
    \hline
    \multirow{6}{7em}{XBRL Dataset (\textbf{1700})} 
                            & XBRL Term & 500  & FActScore &  XBRL Agent\\
                            & Domain Query & 50 & FActScore & XBRL Agent \\
                            & Financial Math & 1000 & Accuracy & XBRL Agent\\
                            & Numeric Query & 50 & FActScore & XBRL Agent \\
                            & Tag Query & 50 & Accuracy & XBRL filings from SEC\\
                            & Financial Ratio Formulas & 50 & Accuracy & XBRL filings from SEC\\
    \hline
  \end{tabular}
  \caption{Statistics of datasets with domains, size, evaluation metrics, and data sources.}
  \label{tab:datasets}
\end{table*}

\subsection{Basic Capabilities (Task 1-5)}
To assess LLMs' basic capabilities in financial information retrieval, we design five basic tasks. As shown in Table \ref{tab:tasks}, the tasks are as follows: 
\begin{itemize}[leftmargin=*]
    \item \textbf{Abbreviation Recognition}. Recognize stock tickers and acronyms for regulation terms.
    \item \textbf{Definition Recognition}. Retrieve the definitions of terms and phrases to ensure compliance.
    \item \textbf{Named Entity Recognition (NER)}. Identify entities such as organizations, legislation, dates, addresses, monetary value, and statistics.
    \item \textbf{Question Answering}. Answer questions regarding given regulatory documents.
    \item \textbf{Link Retrieval}. Retrieve and provide links to particular regulations.
\end{itemize}
We identify important sectors and regulatory agencies, including the OTC derivative market regulated under the European Market Infrastructure Regulation (EMIR), the U.S. securities market regulated by the U.S. Securities and Exchange Commission (SEC), the U.S. banking system primarily overseen by the Federal Reserve, and Generally Accepted Accounting Principles (GAAP), which provide accounting and auditing standards.

\textbf{Question Sets}. We create question sets based on glossaries, FAQs, handbooks, and regulations from official websites.

\subsection{Passing Certificate (Task 6)}
\textbf{Task Description}. This task aims to assess LLMs' ability to accurately answer certificate-level questions about ethics and regulations. The questions are sourced from the three levels of the Chartered Financial Analyst (CFA) exams and the Regulation (REG) section of the Certified Public Accountant (CPA) exam. Both exams cover a wide range of practice scenarios in finance and accounting, which are essential for compliance with applicable legal and ethical standards.

\textbf{Question Set}. This question set includes multiple-choice questions from all three levels of CFA mock/real exams, as well as REG CPA mock exams. Each CFA question has three answer choices. Some questions are grouped to share a common context. Each CPA REG question has four answer choices.

\textbf{Disclaimer: This question set is stored privately and will not be released. They are only used for research purposes internally. We do not and will not share any questions with external researchers.}

\subsection{Common Domain Model (Task 7)}
\textbf{Task Description}. In this task, we assess LLMs' ability to answer questions related to the Common Domain Model (CDM)\footnote{Website of CDM at FINOS: \url{https://cdm.finos.org/}}. CDM is a machine-oriented model for managing the lifecycle of financial products and transactions. It aims to enhance the efficiency and regulatory oversight of financial markets. For this new machine-oriented standard, LLMs can help the financial community understand CDM's modeling approach, use cases, and deployment, thereby enhancing its promotion.

\textbf{Question Set}. The CDM question set comprises a collection of questions and answers derived from the CDM documentation. As shown in Table \ref{tab:datasets}, we generate 80 question-answer pairs about basic definitions and concepts across 5 modeling dimensions, including the product model, event model, legal agreements, process model, and other general aspects. We also collect 46 questions about model implementation and deployment, provided by FAQs and experts at FINOS, Linux Foundation.

\subsection{MOF Licenses (Task 8)}
\textbf{Task Description}. In this task, we assess LLMs' ability to answer questions about the licensing requirements outlined in the MOF ~\cite{white2024mof}. The MOF evaluates and classifies the completeness and openness of machine learning models. The MOF decomposes models into 17 components, each with specific licensing requirements to ensure openness. LLMs can help the open source community better understand the requirements for model openness and avoid misleading openwashing behaviors. 

\textbf{Question Set}. The question set includes license abbreviations, yes/no questions about whether the Open Source Initiative (OSI) approves licenses, and questions about license requirements outlined in the MOF. Expanding the abbreviations of OSI-approved licenses\footnote{The MOF framework encourages OSI-approved licenses: \url{https://opensource.org/licenses}} and judging OSI approval are essential capabilities for classifying model openness. In addition, we also create question-and-answer pairs about model components and their licensing requirements under the MOF. 

\subsection{XBRL Analytics (Task 9)}
\textbf{Task Description}. This task aims to assess LLMs' ability to retrieve and interpret XBRL filings. XBRL is a standard for electronic communication of business and financial data ~\cite{han2024xbrl}. The SEC mandates the submission of XBRL filings for financial statements, but there is a high error rate in the filing process. LLMs can help industries and companies prepare and verify XBRL filings to reduce errors. 

\textbf{Question Set}. 
We utilize the dataset developed by XBRL Agent ~\cite{han2024xbrl} to test LLMs' ability to explain XBRL terms, answer domain and numeric questions based on XBRL reports, and perform financial math calculations. In addition, to better evaluate LLMs' ability to recognize and apply tags in XBRL filings, we create 50 tag queries that ask for the specific tag for a financial item in basic financial statements and 50 questions about financial ratio formulas that ask for the formula written with corresponding tags. Five years of XBRL filings of Dow Jones 30 companies are obtained from the SEC website.

\section{Participants}
There were 25 teams registered for the Regulations Challenge, out of which 6 teams submitted their full solutions.  
We specify three baseline models: Llama 3.1-8B \cite{llama3.1}, GPT-4o \cite{hurst2024gpt}, and Mistral Large 2 \cite{mistral_large_2}. GPT-4o and Mistral Large 2 are selected for their strong performance, while Llama 3.1-8B is chosen because its model size is manageable for participants. Some teams' methods are as follows:

\begin{itemize} [leftmargin=*]
    \item \textbf{FinMind-Y-Me} \cite{finmind2024regchallenge} fine-tuned the Qwen 2.5-7B-Instruct model using sequential fine-tuning, reasoning-based training, and Chain-of-Thought (CoT) inferencing. FinMind-Y-Me’s model is the top-performing model in the Regulations Challenge.
    \item \textbf{IntelliChain Stars} \cite{intellichain2024regchallenge} used a dataset with 30,000 samples of proprietary financial regulations and general financial texts, processed through a pipeline with semantic screening, quality filtering, and deduplication. They used this dataset to fine-tune Llama 3.2-3B-Instruct ~\cite{llama3.2}. 
    \item \textbf{Uniandes} \cite{uniandes2024regchallenge} employed continual pretraining of the Llama 3.1-8B model using a corpus of financial and regulatory documents and then fine-tuned the model using Quantized Low-Rank Adaptation (QLoRA) \cite{dettmers2024qlora} across all nine tasks.
    \item \textbf{Audit-FT} \cite{auditft2024regchallenge} fine-tuned the Qwen 7B-chat \cite{bai2023qwen} model using the Audit Instruction Tuning dataset. This dataset consists of 15 audit tasks across sentence, paragraph, and document levels, such as relation classification, audit issue summary, and document generation. 
\end{itemize}

\begin{sidewaystable}
\centering
    \begin{tabular}{c l p{2cm}| p{2cm} | p{1.5cm} | p{1cm} | p{1cm} | p{1.5cm} |p{1cm} p{1cm} p{1cm} p{1cm} p{1cm}}
        \hline
        \textbf{Ranking} & \textbf{Team Name} & \textbf{Final Score (Weighted)} & \textbf{Abbreviation} & \textbf{Definition} & \textbf{NER} & \textbf{QA} & \textbf{Link Retrieval} & \multicolumn{5}{c}{\textbf{Certificate}}\\
        & & & & & & & & Avg. & CFA I & CFA II & CFA III & REG CPA \\
        \hline
        1 & FinMind-Y-Me & 0.54801 & 0.2095 & 0.5849 & 0.7174 & 0.8609 & 0.2360 & 0.4701 & 0.4889 & 0.4675 & 0.4487 & 0.4752\\
        2 & Uniandes & 0.43929 & 0.2748 & 0.4688 & 0.4302 & 0.7688 & 0.0435 & 0.3112 & 0.3444 & 0.2857 & 0.3077 & 0.3069 \\
        3 & GGBond & 0.43798 & 0.1959 & 0.3800 & 0.6268 & 0.6181 & 0.0621 & 0.3700 & 0.4222 & 0.3506 & 0.4103 & 0.2970\\
        4 & Audit-FT & 0.36075 & 0.1464 & 0.5359 & 0.0000 & 0.6596 & 0.0062 & 0.4020 & 0.4667 & 0.4286 & 0.3462 & 0.3663\\
        5 & IntelliChain Stars & 0.34017 & 0.0698 & 0.4505 & 0.0000 & 0.5628 & 0.0000  & 0.4235 & 0.4778 & 0.3506 & 0.4103 & 0.4554\\
        6 & finma & 0.32286 & 0.0653 & 0.5112 & 0.0000 & 0.5984 & 0.0000 & 0.3266 & 0.4111 & 0.2987 & 0.3590 & 0.2376\\
        \hline
        Baseline & Llama 3.1-8B & 0.53572 & 0.2320 & 0.5130 & 0.6352 & 0.8079 & 0.4348 & 0.4325 & 0.5111 & 0.4026 & 0.4103 & 0.4059 \\
        Baseline & GPT-4o & 0.63567 & 0.3784 & 0.5520 & 0.7108 & 0.8842 & 0.2050 & 0.6568 & 0.6889 & 0.5714 & 0.6538 & 0.7129\\
        Baseline & Mistral Large 2 & 0.62489 & 0.2230 & 0.5338 & 0.7062 & 0.8263 & 0.5875 & 0.6330 & 0.6889 & 0.5584 & 0.6410 & 0.6436\\
        \hline
    \end{tabular}
\caption{The rankings of teams and evaluation results for Tasks 1-6.}
\label{tab:evaluation1-6}
\end{sidewaystable}

\begin{sidewaystable}
\centering
    \begin{tabular}{c l | c | p{1cm} p{1cm} p{1.3cm} p{2cm} | p{1cm} p{1cm} p{2.3cm} p{2cm} p{1cm} }
        \hline
        \textbf{Ranking} & \textbf{Team Name}  & \textbf{CDM} & \multicolumn{4}{c}{\textbf{MOF Licenses}} & \multicolumn{5}{c}{\textbf{XBRL Analytics}}\\
        &  & &  MOF Avg. & License Abbr. & OSI  Approval & Detailed QA & XBRL Avg. & XBRL Term & Domain \& Numeric Query & Financial Math & Tag Query\\
        \hline
        1 & FinMind-Y-Me & 0.8528 & 0.5266 & 0.0323 & 0.7400 & 0.8075 & 0.5519 & 0.6327 & 0.6636 & 0.6444 & 0.2667 \\
        2 & Uniandes  & 0.7587 & 0.5373 & 0.2258 & 0.6200 & 0.7660 & 0.4885 & 0.7236 & 0.6636 & 0.5000 & 0.0667\\
        3 & GGBond & 0.8006 & 0.4976 & 0.0000 & 0.8000 & 0.6929 & 0.4586 & 0.6870 & 0.5252 & 0.3111 & 0.3111\\
        4 & Audit-FT & 0.7149 & 0.4202 & 0.0645 & 0.6000 & 0.5961 & 0.3204 & 0.7362 & 0.4122 & 0.1333 & 0.0000\\
        5 & IntelliChain Stars & 0.6635 & 0.4412 & 0.0968 & 0.7000 & 0.5267 & 0.3669 & 0.6539 & 0.5248 & 0.2667 & 0.0222\\
        6 & finma & 0.7045 & 0.3862 & 0.0323 & 0.5200 & 0.6063  & 0.3098 & 0.7242 & 0.4149 & 0.0778 & 0.0222 \\
        \hline
        Baseline & Llama 3.1-8B  & 0.7980 & 0.5149 & 0.1290 & 0.7200 & 0.6956 & 0.5556 & 0.7083 & 0.5845 & 0.7667 & 0.1667\\
        Baseline & GPT-4o & 0.8820 & 0.6564 & 0.1935 & 0.9600 & 0.8156  & 0.7743 & 0.8503 & 0.5851 & 0.8842 & 0.7778\\
        Baseline & Mistral Large 2  & 0.8632 & 0.4640 & 0.1290 & 0.4400 & 0.8229 & 0.7791 & 0.8221 & 0.6831 & 0.7444 & 0.8667\\
        \hline
    \end{tabular}
\caption{Evaluation results for Tasks 7-9.}
\label{tab:evaluation7-9}
\end{sidewaystable}

\section{Evaluation and Discussion}
\subsection{Evaluation}
We split our question dataset into a validation dataset ($10\%$) and a testing dataset ($90\%$). Due to time constraints, we randomly sample 200 questions from stock tickers in abbreviation recognition and 90 questions from financial math in XBRL analytics. We also excluded financial ratio formula queries in XBRL analytics. The evaluation metrics include accuracy, F1 score, BertScore ~\cite{zhang2020bertscore}, and FActScore ~\cite{min2023factscore}, as shown in Table \ref{tab:datasets}. The final score is determined by the weighted average of performance across 9 tasks, with a weight of $10\%$ assigned to each of Tasks 1–5, $20\%$ to Task 6, and $10\%$ to each of Tasks 7–9.

\subsection{Results}
The results are shown in Tables \ref{tab:evaluation1-6} and \ref{tab:evaluation7-9}. FinMind-Y-Me achieves the top position with a final score of 0.54801, outperforming Llama 3.1-8B. Uniandes ranks second, followed by GGBond. 

In some tasks, there are significant performance gaps between models. In the NER task, FinMind-Y-Me achieves a score of 0.7174, while three models fail to correctly identify any single entity. In link retrieval, FinMind-Y-Me leads the submitted models with a score of only 0.2360, far below Mistral Large 2’s score of 0.5875. 

In XBRL analytics, FinMind-Y-Me is the best-performing submitted model, achieving an average score of 0.5519. Among the subtasks, all other submitted models perform equally well or better in the XBRL term explanation, but their performances drop for the remaining XBRL tasks. 

In the MOF task, the top submitted model, Uniandes, achieves an average score of 0.5373, surpassing the score of its base model, Llama 3.1-8B. The license abbreviation subtask is challenging for all models, with no models scoring above 0.23. In the OSI license approval and detailed QA subtasks, the submitted models perform relatively well.

\subsection{Discussion}
GPT-4o and Mistral Large 2 outperform the other models, likely because of their larger model sizes compared to the other models, which have about 8 billion parameters. FinMind-Y-Me’s win highlights the effectiveness of reasoning enhancements.

Among the $9$ tasks, all models perform well in question-answering-related tasks, such as the QA, MOF detailed QA, CDM, and XBRL term explanation tasks. It shows that LLMs have enough factual knowledge about these questions. However, all models perform poorly in abbreviation tasks, such as financial term acronyms, stock tickers, and OSI license abbreviations. It reflects LLMs' deficiency in recognizing abbreviations and responding with accurate full names in financial regulations. In link retrieval, the low accuracy of all models indicates that models have difficulties in searching for and locating online documents. In the NER task, the zero score three models received shows that domain-specific entity extraction is challenging for models that are not fine-tuned effectively. 

For the certificate task, the submitted models underperform compared to GPT-4o and Mistral Large 2, likely because of deficiencies in reasoning and knowledge. FinMind-Y-Me employs reasoning-based training and achieves the highest score among contestants. Audit-FT and IntelliChain Starts both use audit datasets for fine-tuning, providing them with sufficient accounting and auditing knowledge. 

In XBRL analytics, the submitted models perform poorly in the financial math and tag query tasks. Uniandes outperforms its base model, Llama 3.1-8B, in the XBRL term and domain and numeric query tasks, but underperforms in the financial math and tag query tasks. This suggests that domain-specific fine-tuning may reduce other capabilities of base LLMs. In addition, integrating an external XBRL filing database by using retrieval-augmented generation (RAG) may improve models' performance in the tag query task.

\section{Conclusion and Future Work}

In the Regulations Challenge, we created nine novel tasks and corresponding question sets to assess LLMs' ability to understand and interpret financial regulations and industry standards, and also LLMs' understanding of financial products and markets. Through it, we encouraged the academic community to identify the strengths and limitations of LLMs in financial regulations and gain insights into their professional readiness. 

We will organize follow-up challenges on financial regulations. The question sets and evaluation results will be merged back to the Open FinLLM Leaderboard on Hugging Face \cite{colin2024leaderboard,xie2024finben}. To better showcase use cases, we will provide demos by leveraging FinGPT Search Agent \cite{liu2023fingpt,tian2024searchagent}.

\section*{Acknowledgement} 

Keyi Wang and Xiao-Yang Liu Yanglet acknowledge the support from Columbia's SIRS and STAR Program, as well as The Tang Family Fund for Research Innovations in FinTech, Engineering, and Business Operations. 
Jaisal Patel, Andy Zhu, Steve Yang, and Xiao-Yang Liu Yanglet acknowledge the support from a NSF IUCRC CRAFT center research grant (CRAFT Grant 22017) for this research. The opinions expressed in this publication do not necessarily represent the views of NSF IUCRC CRAFT.

\bibliography{custom}

\begin{thebibliography}{24}
\providecommand{\natexlab}[1]{#1}

\bibitem[{Bai et~al.(2023)Bai, Bai, Chu, Cui, Dang, Deng, Fan, Ge, Han, Huang et~al.}]{bai2023qwen}
Jinze Bai, Shuai Bai, Yunfei Chu, Zeyu Cui, Kai Dang, Xiaodong Deng, Yang Fan, Wenbin Ge, Yu~Han, Fei Huang, et~al. 2023.
\newblock Qwen technical report.
\newblock \emph{arXiv preprint arXiv:2309.16609}.

\bibitem[{Brunnermeier et~al.(2009)Brunnermeier, Crockett, Goodhart, Persaud, and Shin}]{markus2009regulations}
Markus Brunnermeier, Andrew Crockett, Charles Goodhart, Avi Persaud, and Hyun Shin. 2009.
\newblock \emph{The fundamental principles of financial regulation}.
\newblock International Center for Monetary and Banking Studies Centre for Economic Policy Research, Geneva London.

\bibitem[{Carrión et~al.(2024)Carrión, Castañeda, and Manrique}]{uniandes2024regchallenge}
Santiago~Martínez Carrión, Juan~Manuel Castañeda, and Rubén Manrique. 2024.
\newblock Uniandes at the regulations challenge task: A scalable framework for legal text understanding in regulatory and financial contexts.
\newblock In \emph{Proceedings of the Joint Workshop of the 9th Financial Technology and Natural Language Processing (FinNLP), the 6th Financial Narrative Processing (FNP), and the 1st Workshop on Large Language Models for Finance and Legal (LLMFinLegal)}.

\bibitem[{Chantangphol et~al.(2024)Chantangphol, Balee, Sucharitpongpan, Saetia, and Chalothorn}]{finmind2024regchallenge}
Pantid Chantangphol, Pornchanan Balee, Kantapong Sucharitpongpan, Chanatip Saetia, and Tawunrat Chalothorn. 2024.
\newblock Finmind-y-me at the regulations challenge task: Financial mind your meaning based on thalle.
\newblock In \emph{Proceedings of the Joint Workshop of the 9th Financial Technology and Natural Language Processing (FinNLP), the 6th Financial Narrative Processing (FNP), and the 1st Workshop on Large Language Models for Finance and Legal (LLMFinLegal)}.

\bibitem[{Dettmers et~al.(2024)Dettmers, Pagnoni, Holtzman, and Zettlemoyer}]{dettmers2024qlora}
Tim Dettmers, Artidoro Pagnoni, Ari Holtzman, and Luke Zettlemoyer. 2024.
\newblock Qlora: Efficient finetuning of quantized llms.
\newblock \emph{Advances in Neural Information Processing Systems}, 36.

\bibitem[{Han et~al.(2024)Han, Kang, Jin, Liu, and Yang}]{han2024xbrl}
Shijie Han, Haoqiang Kang, Bo~Jin, Xiao-Yang Liu, and Steve Yang. 2024.
\newblock {XBRL Agent}: Leveraging large language models for financial report analysis.
\newblock In \emph{ACM International Conference on AI in Finance}.

\bibitem[{Huang et~al.()Huang, Jiang, and Zhu}]{auditft2024regchallenge}
Jiajia Huang, Maowei Jiang, and Haoran Zhu.
\newblock Audit-ft at the regulations challenge task: An open-source large language model for audit.
\newblock In \emph{Proceedings of the Joint Workshop of the 9th Financial Technology and Natural Language Processing (FinNLP), the 6th Financial Narrative Processing (FNP), and the 1st Workshop on Large Language Models for Finance and Legal (LLMFinLegal)}.

\bibitem[{Hurst et~al.(2024)Hurst, Lerer, Goucher, Perelman, Ramesh, Clark, Ostrow, Welihinda, Hayes, Radford et~al.}]{hurst2024gpt}
Aaron Hurst, Adam Lerer, Adam~P Goucher, Adam Perelman, Aditya Ramesh, Aidan Clark, AJ~Ostrow, Akila Welihinda, Alan Hayes, Alec Radford, et~al. 2024.
\newblock Gpt-4o system card.
\newblock \emph{arXiv preprint arXiv:2410.21276}.

\bibitem[{Jiang et~al.(2024)Jiang, Dai, Jia, Wang, and Wang}]{intellichain2024regchallenge}
Shijia Jiang, Yongfu Dai, Haochen Jia, Yuxin Wang, and Hao Wang. 2024.
\newblock Intellichain stars at the regulations challenge task: A large language model for financial regulation.
\newblock In \emph{Proceedings of the Joint Workshop of the 9th Financial Technology and Natural Language Processing (FinNLP), the 6th Financial Narrative Processing (FNP), and the 1st Workshop on Large Language Models for Finance and Legal (LLMFinLegal)}.

\bibitem[{Kang and Liu(2023)}]{kang2023deficiency}
Haoqiang Kang and Xiao-Yang Liu. 2023.
\newblock Deficiency of large language models in finance: An empirical examination of hallucination.
\newblock In \emph{I Can't Believe It's Not Better Workshop: Failure Modes in the Age of Foundation Models (NeurIPS)}.

\bibitem[{Labonte(2023)}]{labonte2023framework}
Marc Labonte. 2023.
\newblock Who regulates whom? an overview of the {U.S.} financial regulatory framework.
\newblock \emph{Congressional Research Service Report}.

\bibitem[{Lin et~al.(2024)Lin, Tian, Wang, Zhao, Huang, Xie, Borella, White, Wang, Xiao, Yanglet, and Deng}]{colin2024leaderboard}
Shengyuan~Colin Lin, Felix Tian, Keyi Wang, Xingjian Zhao, Jimin Huang, Qianqian Xie, Luca Borella, Matt White, Christina~Dan Wang, Kairong Xiao, Xiao-Yang~Liu Yanglet, and Li~Deng. 2024.
\newblock Open {FinLLM} leaderboard: Towards financial ai readiness.
\newblock \emph{International Workshop on Multimodal Financial Foundation Models (MFFMs) at 5th ACM International Conference on AI in Finance (MFFM at ICAIF ’24)}.

\bibitem[{Liu et~al.(2023)Liu, Wang, Yang, and Zha}]{liu2023fingpt}
Xiao-Yang Liu, Guoxuan Wang, Hongyang Yang, and Daochen Zha. 2023.
\newblock Data-centric fingpt: Democratizing internet-scale data for financial large language models.
\newblock In \emph{Workshop on Instruction Tuning and Instruction Following, NeurIPS}.

\bibitem[{{Meta AI}(2024{\natexlab{a}})}]{llama3.1}
{Meta AI}. 2024{\natexlab{a}}.
\newblock \href {https://ai.meta.com/research/publications/the-llama-3-herd-of-models/} {The llama 3 herd of models}.

\bibitem[{{Meta AI}(2024{\natexlab{b}})}]{llama3.2}
{Meta AI}. 2024{\natexlab{b}}.
\newblock \href {https://ai.meta.com/blog/llama-3-2-connect-2024-vision-edge-mobile-devices/} {Llama 3.2: Revolutionizing edge {AI} and vision with open, customizable models}.

\bibitem[{Min et~al.(2023)Min, Krishna, Lyu, Lewis, tau Yih, Koh, Iyyer, Zettlemoyer, and Hajishirzi}]{min2023factscore}
Sewon Min, Kalpesh Krishna, Xinxi Lyu, Mike Lewis, Wen tau Yih, Pang~Wei Koh, Mohit Iyyer, Luke Zettlemoyer, and Hannaneh Hajishirzi. 2023.
\newblock \href {https://arxiv.org/abs/2305.14251} {Factscore: Fine-grained atomic evaluation of factual precision in long form text generation}.
\newblock \emph{Preprint}, arXiv:2305.14251.

\bibitem[{{Mistral AI}(2024)}]{mistral_large_2}
{Mistral AI}. 2024.
\newblock \href {https://mistral.ai/news/mistral-large-2407/} {Large enough}.

\bibitem[{Osborne et~al.(2024)Osborne, Ding, and Kirk}]{osborne2024ai}
Cailean Osborne, Jennifer Ding, and Hannah~Rose Kirk. 2024.
\newblock {The {AI} community building the future? A quantitative analysis of development activity on Hugging Face Hub}.
\newblock \emph{Journal of Computational Social Science}, pages 1--39.

\bibitem[{Roberts et~al.(2023)Roberts, Ziosi, Osborne, Saouma, Belias, Buchser, Casovan, Kerry, Meltzer, Mohit, Ouimette, Renda, Stix, Teather, Woodhouse, and Zeng}]{roberts2023}
H.~Roberts, M.~Ziosi, C.~Osborne, L.~Saouma, A.~Belias, M.~Buchser, A.~Casovan, C.~Kerry, J.~Meltzer, S.~Mohit, M.-E. Ouimette, A.~Renda, C.~Stix, E.~Teather, R.~Woodhouse, and Y.~Zeng. 2023.
\newblock \href {https://ceimia.org/wp-content/uploads/2023/02/Comparative-Framework-for-AI-Regulatory-Policy.pdf} {A comparative framework for {AI} regulatory policy}.
\newblock CEIMIA.

\bibitem[{Tian et~al.(2024)Tian, Byadgi, Kim, Zha, White, Xiao, and Liu}]{tian2024searchagent}
Felix Tian, Ajay Byadgi, Daniel~S Kim, Daochen Zha, Matt White, Kairong Xiao, and Xiao-Yang Liu. 2024.
\newblock Customized fingpt search agents using foundation models.
\newblock In \emph{ACM International Conference on AI in Finance}.

\bibitem[{White et~al.(2024)White, Haddad, Osborne, Yanglet, Abdelmonsef, and Varghese}]{white2024mof}
Matt White, Ibrahim Haddad, Cailean Osborne, Xiao-Yang~Liu Yanglet, Ahmed Abdelmonsef, and Sachin Varghese. 2024.
\newblock \href {https://arxiv.org/abs/2403.13784} {The model openness framework: Promoting completeness and openness for reproducibility, transparency, and usability in artificial intelligence}.
\newblock \emph{Preprint}, arXiv:2403.13784.

\bibitem[{Xie et~al.(2024)Xie, Han, Chen, Xiang, Zhang, He, Xiao, Li, Dai, Feng, Xu, Kang, Kuang, Yuan, Yang, Luo, Zhang, Liu, Xiong, Deng, Jiang, Yao, Li, Yu, Hu, Huang, Liu, Lopez-Lira, Wang, Lai, Wang, Peng, Ananiadou, and Huang}]{xie2024finben}
Qianqian Xie, Weiguang Han, Zhengyu Chen, Ruoyu Xiang, Xiao Zhang, Yueru He, Mengxi Xiao, Dong Li, Yongfu Dai, Duanyu Feng, Yijing Xu, Haoqiang Kang, Ziyan Kuang, Chenhan Yuan, Kailai Yang, Zheheng Luo, Tianlin Zhang, Zhiwei Liu, Guojun Xiong, Zhiyang Deng, Yuechen Jiang, Zhiyuan Yao, Haohang Li, Yangyang Yu, Gang Hu, Jiajia Huang, Xiao-Yang Liu, Alejandro Lopez-Lira, Benyou Wang, Yanzhao Lai, Hao Wang, Min Peng, Sophia Ananiadou, and Jimin Huang. 2024.
\newblock Finben: A holistic financial benchmark for large language models.
\newblock \emph{NeurIPS, Special Track on Datasets and Benchmarks}.

\bibitem[{Yanglet and Deng(2024)}]{liu2024MFFM}
Xiao-Yang~Liu Yanglet and Li~Deng. 2024.
\newblock Multimodal financial foundation models (mffms): Progress, prospects, and challenges.
\newblock \emph{International Workshop on Multimodal Financial Foundation Models (MFFMs) at 5th ACM International Conference on AI in Finance (MFFM at ICAIF ’24),}.

\bibitem[{Zhang et~al.(2023)Zhang, Kishore, Wu, Weinberger, and Artzi}]{zhang2020bertscore}
Tianyi Zhang, Varsha Kishore, Felix Wu, Kilian~Q. Weinberger, and Yoav Artzi. 2023.
\newblock {BERTScore}: Evaluating text generation with bert.
\newblock In \emph{International Conference on Learning Representations}.

\end{thebibliography}

\end{document}